# Towards the next generation of exergames: Flexible and personalised assessment-based identification of tennis swings

Boris Bačić
School of Engineering, Computing & Mathematical Sciences
Auckland University of Technology
Auckland, New Zealand
boris.bacic@aut.ac.nz

*Abstract*—Current exergaming sensors and inertial systems attached to sports equipment or the human body can provide quantitative information about the movement or impact e.g. with the ball. However, the scope of these technologies is not to qualitatively assess sports technique at a personalised level, similar to a coach during training or replay analysis. The aim of this paper is to demonstrate a novel approach to automate identification of tennis swings executed with erroneous technique without recorded ball impact. The presented spatiotemporal transformations relying on motion gradient vector flow and polynomial regression with RBF classifier, can identify previously unseen erroneous swings (84.5–94.6%). The presented solution is able to learn from a small dataset and capture two subjective swing-technique assessment criteria from a coach. Personalised and flexible assessment criteria required for players of diverse skill levels and various coaching scenarios were demonstrated by assigning different labelling criteria for identifying similar spatiotemporal patterns of tennis swings.

I. INTRODUCTION

Current wearables and sports equipment with inertial sensors can detect specific motion patterns and provide a range of quantitative analyses, but such technology cannot teach end-users how to improve punch, kick or (golf) swing technique [1, 2]. For an exergame to emulate a broadcasted TV sport event experience, it would be a desired feature for the participant(s) to see the replay of a good or bad movement with running commentary similar to a broadcaster's expert panel providing subjective opinions, and strategic and coaching advice. For augmented coaching and rehabilitation monitoring system design, it would be a desired feature to capture personalised assessments emulating a physiotherapist assisting and monitoring an athlete's progress with injury/sport-specific exercises before returning to the sport. The systems and technology that could quantify qualitative assessment of human movement would be applicable to several fields such as exergames, technology-mediated coaching practice and team selection, as well as activity, health and rehabilitation monitoring technologies.

To design the next generation of exergames and augmented coaching systems and technology (ACST), a common obstacle is to find a solution for how to distinguish between good or bad movement patterns based on qualitative and subjective criteria as opposed to quantitative criteria relying on measured results of the movement. Inspired to overcome this obstacle, this study provides an investigation for the related questions: (1) Can a machine detect good and bad tennis swing technique? (2) Is it possible for a machine to capture a subjective expert's swing technique assessment from replay using a small training dataset? and (3) Can anonymised replay such as a 3D animated stick figure be used for expert assessment of tennis swings?

Early investigation of the *swing plane* concept in golf [2] demonstrates that it is possible to quantify common-sense descriptive rules that guide coaching feedback and provide validity using a data-driven AI approach for such rules. Given that there is little work available on the use of AI and specifically computational intelligence for *human motion modelling and analysis* (HMMA), this multi-disciplinary work aims to contribute to *computational sport science* and advancements of the next generation of exergames and ACST.

II. BACKGROUND

*A. Exergames, Augmented Coaching Systems and eSports*

Since the inception of exergaming consoles with Nintendo Wii and later with Microsoft Kinect, it seems that there have been no substantial advancements in this particular genre when compared to the rising popularity of video games included in international eSports tournaments (e.g. www.espn.com/esports/ and http://dailyesports.tv/). While both versions of Kinect sensors provide unobtrusive marker-less depth data acquisition streamed at 30 fps and are considered by the scientific community as an open-source hardware with Windows-based proprietary SDK and open-source software libraries, exergames are not considered as part of the eSports community, which according to a CNN projection will grow in revenue to $1 billion by 2019 [3]. In spite of the health benefits of physical movement, there is no obvious inclusion of exergaming in eSports. The possible reasons for the exclusion of exergaming in eSports are: privacy preservation concerns and a lack of competitive, strategic team play (including sense of belonging and level of emotions), challenges associated with player's movement goals (including reaction times and proprioceptive feel) and perceived 'fairness' of movement assessment.

In the physical world, the ball flight as an outcome of a swing, represents quantitative *knowledge of results* (KR), which in exergaming is typically not recorded. While the advancements in commercial golf immersive reality applications may be the best candidate for inclusion in eSports, there is still an open question of the 'fairness' of movement assessment for *approach shot* and judging the *putting* trajectory for each green. Knowing that in gymnastics, ice skating, and other stylistic-execution sports there is a panel of judges assigned to evaluate performance it seems obvious that qualitative evaluation based on hard-to-define assessment criteria is a significant challenge for AI. One of the first scientific investigations on the effects of augmented coaching feedback on elements of performance was reported in 1976 [4]. From Hatze's research in biomechanics [4], it is possible to generalise that if training for a given motor learning task is based only on KR feedback, the participant's performance will plateau due to his/her natural adaptation. However, it is also possible to further improve the subject's performance by providing qualitative feedback that is based on *knowledge of performance* (KP) i.e. knowledge about the elements of performance associated with the goal(s) of the movement.

### B. Motion Acquisition Technology, Privacy and Data Ownership

Modern inertial systems attached to sports equipment and the human body use proprietary algorithms to report quantitative data. Unfortunately, raising legal concerns [5, 6], such systems do not provide end-users with option to own the recorded raw data on expressed motion patterns, which are typically processed on third party's cloud and are not shared with the broader scientific community. In tennis, for example various sensors attached to the racquet's handle (http://en.babolatplay.com, www.smarttennissensor.sony.net, and www.zepp.com), can provide training statistics including the number of particular shots hit, the estimated ball rotation, speed at point of impact and even whether the player has missed the optimal impact zone of the racquet's string bed. Although such information can be used for swing-quality assessment or to produce other similar assessments based on quantitative criteria, mishitting the ball is still possible regardless of a good or bad swing technique and no available systems so far are able to index swings based on swing technique that would be based on qualitative personalised and flexible assessment criteria.

### C. Tennis, Coaching, and Sports Analysis: Neural Signal Processing Perspective

Tennis is considered an open-skill sport where opponents influence each other's choices. Stylistic execution of tennis swings is subject to skill-level, game situation (e.g. defending or attacking), and other personal idiosyncrasies. Unlike coaching at an elite-level, when coaching beginners, a coach is expected to recognise, prioritise, produce feedback and recommend intervention for *common errors* that beginners typically show during their play. Furthermore, in tennis, it is considered as general knowledge that: (1) racquet and string technology have evolved since the times of wooden racquets and (2) advanced-level players are sensitive about the subtle differences between seemingly identical racquets and about when to swap the racquets during the match.

At an advanced and professional competitive level, tennis players often describe their state of mind as *being in the zone* and racquet interaction with the ball as *feel*. The area on the string bed surface that has the best feel is known as the *sweet spot* and is typically included in a racquet's specification.

> "A coach can teach many things, but they cannot teach *feel*. That is something you must master on your own."
> [*Nick Bolletieri, tennis coach*]

How the sweet spot influences racquet choice and the racquet's feel and what happens in the brain of an experienced player when swinging the racquet through the air outside of the tennis court may be interesting questions for neuroscience, but what is pertinent to computer science is the question of whether we can model and emulate this feel.

## III. METHODOLOGY

The 3D motion dataset utilised in this study was recorded in a biomechanics laboratory using eMotion (BTS) SMART-e 900, a nine-camera optoelectronic motion system. The capture volume where the tennis swings were recorded was approximately 3x2x2 m. The utilised minimalistic retro-reflective marker set to produce a 3D stick figure (Fig. 1) is similar to Kinect sensor, but with additional markers attached to a tennis racquet.

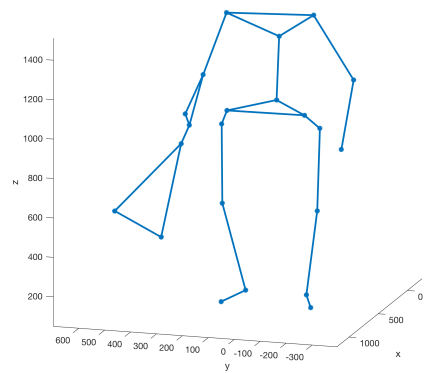

Fig. 1. Stick figure model produced from a minimalistic retro-reflective marker set attached to the tennis racquet and human body.

Compared to MS Kinect™ sensors capturing depth video at 30 Hz, the captured dataset was recorded at 50 Hz with high (sub-millimetre) 3D resolution and was also able to produce additional information about the racquet movement and forearm internal and external rotation as shown in Fig. 1. The tennis swing experimental dataset contains common errors typical for novice to intermediate-level tennis players. The set of forehands were executed as a mix of fast and slow swings from diverse stances (Fig. 2). Due to the variety of forehands, the selected action zones' temporal region of interest (ROI) are not all of the same duration, but last between 7 and 13 frames.

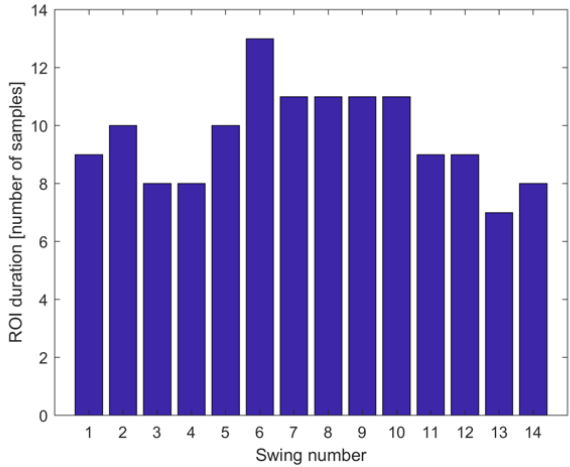

Fig. 2. Action zone durations in experimental dataset.

### A. Design Decisions, Insights and Rationale of the Study

As a design decision, only captured motion data was used in processing without any synthetic data derived from the experimental dataset. For an expert to consistently select a swing's action zone ROI that is 0.14 – 0.26 seconds in duration, a stand-alone 3D stick-figure player [7] was used. The 3D stick-figure player allows pixel-accurate interactive replays using virtual camera 360° view with panning, zooming and other features such as selected A-B sequence replay and variable slow-motion. Accurate and anonymised 3D replay capabilities are considered one of the key tools for visual analysis and human motion modelling and analysis.

The utilised dataset size was considered sufficient for the intended purposes of: (i) feature extraction algorithm development; (ii) model design; (iii) supervised machine learning experiments relying on flexible assessment criteria based on observed *common errors* and their similarity-based grouping; and (iv) small expert-based training data requirement.

To quantify the internalised phenomenon of 'feel' through the impact and action zone, *feature extraction technique* (FET) is represented in this work as a single gradient vector flow of the racquet's sweet spot motion. The FET approach is intended to capture in data the previous state of the racquet and the change in direction and velocity.

In the area of machine learning and neural information processing, the use of a gradient function is typically associated with the well-known gradient descent algorithm. Applied to this study, HMMA and computational sport science in general, gradient function allows: (i) visualisation of two- and three-dimensional curvatures (as contours and vector fields) pointing out the direction of highest changes in space as the gradient vector field where the vector magnitude is associated with the steepness of the slope at a particular point; (ii) computing displacement of a point or a position of a virtual marker in 3D plane; (iii) representing the directional derivative of a function in the direction of computed unit vector(s); (iv) kinematic motion data processing; (v) linear approximation of a function value at a given point; and (vi) mathematical transformations of temporal and spatial region of interest into spatial feature (or spatial pattern) space that can be further transformed and processed by a machine.

For a function *F* of three variables (*x,y,z*) in 3D Cartesian orthogonal space with Euclidean metrics, gradient *F* (1) is denoted as:

$$grad\ F(x,y,z) = \nabla F(x,y,z) = \left(\frac{dF}{dx}, \frac{dF}{dy}, \frac{dF}{dz}\right) \quad (1)$$

Similarly, in biomechanics and computational sport science, it is common to use the notations (2) for kinematic data processing and analysis:

$$\begin{aligned} Position &= (x, y, z); \\ Velocity &= (\dot{x}, \dot{y}, \dot{z}); \\ Acceleration &= (\ddot{x}, \ddot{y}, \ddot{z}); \\ \dot{x} &= \frac{dx}{dt}\ or\ \dot{x} = \frac{\Delta x}{\Delta t}. \end{aligned} \quad (2)$$

The gradient derivative of a function $\nabla F$ (3) in the direction of $\hat{\imath}, \hat{\jmath}, \hat{k}$ unit vectors relative to *x,y,z* coordinate is computed as:

$$\nabla F = \frac{\partial F}{\partial x}\hat{\imath} + \frac{\partial F}{\partial x}\hat{\jmath} + \frac{\partial F}{\partial x}\hat{k} \quad (3)$$

Gradient vector (4) can also be used for changing relative position of a virtual marker on the given plane – as a linear function approximation at the point $x_0$:

$$f(x) \approx f(x_0) + (\nabla f)_{x_0} \cdot (x - x_0) \quad (4)$$

Where the gradient vector at the specific 3D point *p(x,y,z)* (5) is noted as:

$$(\nabla f)_{p(x,y,z)} \quad (5)$$

Considering the applications of the gradient function for motion data processing, the idea of producing a representation of a tennis swing or other sport-specific movement patterns is based on combining the state or position of the initial point $x_0$ and its subsequent states (5) with the resulting gradient motion vector field (3).

## B. Data Pre-Processing

For modelling and analytical purposes using external software, the captured 3D motion dataset $\mathbb{R}^3$ was exported as an ASCII file containing right-handed XYZ marker locations in columns and the samples organised as rows. For visualisation, modelling and analysis in Mathwork's Matlab™, the right-handed XYZ motion dataset was converted into Matlab's default left-handed XZY internal data format, noted as $f: \mathbb{R}^3 \rightarrow \mathbb{R}^3$, where $\langle x_i, y_i, z_i \rangle \triangleq \langle x_i, z_i, -y_i \rangle$.

The stick figure (Fig. 1) is considered as a set $M$ of interconnected markers ($n=22$). Each marker $m_i \in \mathbb{R}^3$ is comprised of the three ($x_i, y_i, z_i$) time series (6) sampled at regular time intervals.

$$m_i = (x_i, y_i, z_i),$$
$$m_i \in M, \; i = 1, ..., n, \quad (6)$$
$$M = (m_1, m_2, ..., m_n).$$

Intended ball impact is a subset of swing action zone ROI.

Visualisation and expert labelling (Fig. 3) show examples of good swings and common errors using overlaid stick figure frames. The racquet path and the estimated sweet spot trajectories are shown as swing volume through action zone – the temporal region of interest. Subjective decision boundaries are depicted with diverse expert assessment decisions that were associated with flexible skill-level criteria of the same swing (Fig. 3b). Furthermore (as in Fig. 3), some bad swings contain multiple issues that a coach would need to prioritise in his/her feedback.

approach and expert insight. Given the large number of forehand variations, including 'good' and 'bad' technique and problem space dimensionality, the reduction of redundant data was based on expert insight. The insight and rationale here is that swing technique is linked to cognitive activity associated with the racquet's feel, which is also linked to proprioception of balance, movement fluidity and timely swing execution as a response to the opponent's activity. The full set of 22 markers of a swing $S_j$ were used for expert visual assessment or swing labelling using an animated 3D stick figure. Selecting a subset of markers on empirical basis is considered as feature reduction and it is also aligned with sports technology, where an inertial sensor is attached to sports equipment. The chosen representation of a sensor would represent the racquet's sweet spot movement. While attaching a marker to the racquet's sweet spot (or other sports equipment in general e.g. golf club) would be obtrusive and impractical, the workaround was to compute a virtual marker (Table I – Algorithm 1). To compute the racquet's 'feel' at impact and through the action zone and produce related discriminative feature set for machine learning purpose, the swing motion data are represented as single motion gradient flow of the racquet's sweet spot. To produce a sweet spot virtual marker's data, a minimum of three markers were needed to compute the racquet plane which must be aligned with the racquet's string bed. The locations of the virtual sweet-spot's marker are to be combined with the spatiotemporal pattern of a computed motion gradient vector flow comprising displacement and changes in marker velocity. Information about location, direction and change in displacement magnitude at regular time intervals is visualised as a vector flow, which was transformed

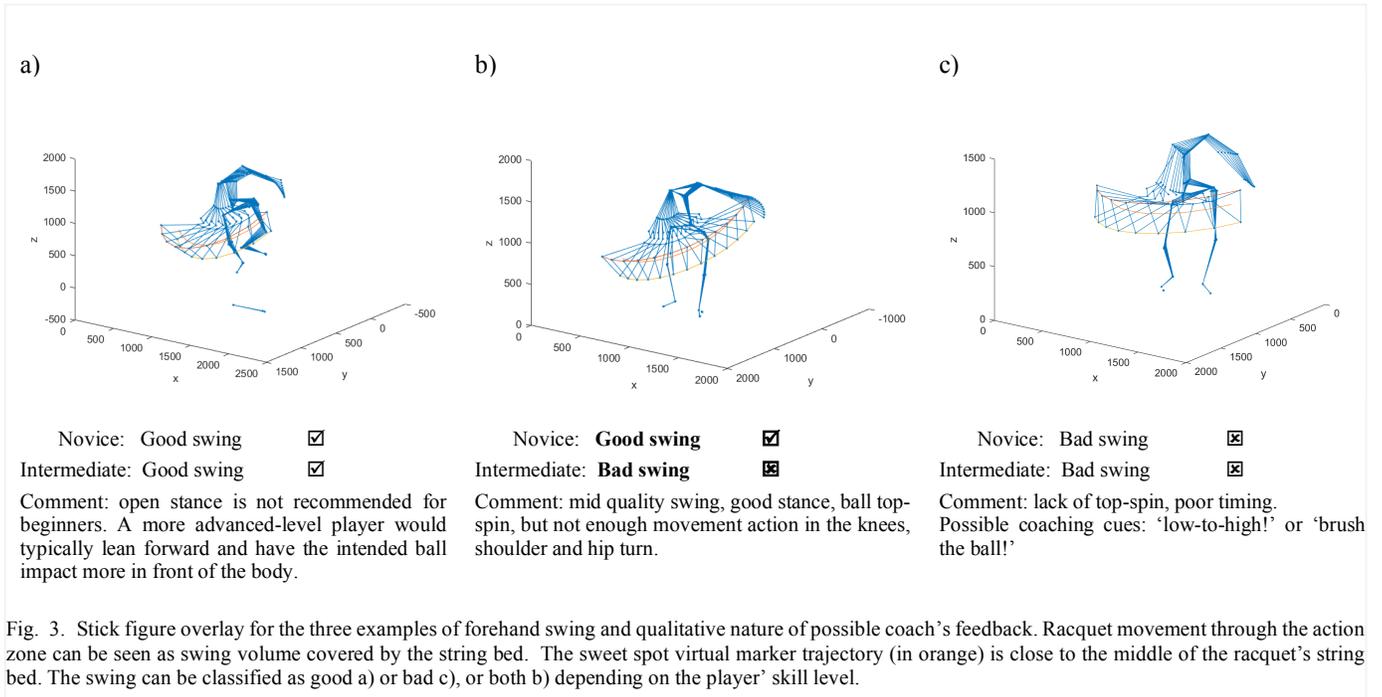

Fig. 3. Stick figure overlay for the three examples of forehand swing and qualitative nature of possible coach's feedback. Racquet movement through the action zone can be seen as swing volume covered by the string bed. The sweet spot virtual marker trajectory (in orange) is close to the middle of the racquet's string bed. The swing can be classified as good a) or bad c), or both b) depending on the player' skill level.

## C. Data Transformation and Feature Extraction

Body and racquet movement are represented by 22 markers resulting in 66 time-series data. Problem space and dimensionality reduction was partially guided by an empirical and expressed as a spatial pattern at later processing stage. The spatiotemporal patterns are computed from the previous state and movement vector flow of the virtual sweet spot. Temporal patterns of motion trajectories and vector flow within the action zone of a swing were projected in *Sagittal* and *Transverse* planes

TABLE I. PSEUDO CODE FOR FEATURE EXTRACTION TECHNIQUE REPRESENTING A SINGLE 3D MARKER MOTION PATTERN OF A TENNIS SWING

**Algorithm 1** Motion Gradient Vector Flow of the Projected Racquet's Sweet Spot

**Require:** Swing's ROI $S_j$: $NaN \cap \{\overrightarrow{M_{R1}}, \overrightarrow{M_{R2}}, \overrightarrow{M_H}\} = \emptyset$
**Ensure:** $f: S_j \rightarrow Y$

{* Compute racquet's 3D sweet spot marker *}[a]
1: $\overrightarrow{M_{SS}} \leftarrow compVirtualMarker(\overrightarrow{M_{R1}}, \overrightarrow{M_{R2}}, \overrightarrow{M_H})$

{* Compute vector array of the sweet spot's movements *}[b]
2: $[\overrightarrow{Ux_{ss}}, \overrightarrow{Vy_{ss}}, \overrightarrow{Wz_{ss}}] \leftarrow grad(\overrightarrow{M_{SS}})$

{* Compute vector array tips as virtual marker *}
3: $[\overrightarrow{Mux_{ss}}, \overrightarrow{Mvy_{ss}}, \overrightarrow{Mwz_{ss}}] \leftarrow compVirtualMarkerTip(\overrightarrow{Ux_{ss}}, \overrightarrow{Vy_{ss}}, \overrightarrow{Wz_{ss}}, \overrightarrow{M_{SS}})$

{* Convert spatiotemporal vectors flow into temporal patterns *}[c]
4: $Sagittal\_Plane \leftarrow [polyFit(\overrightarrow{Mux_{ss}}, \overrightarrow{Mwz_{ss}}), polyFit(\overrightarrow{Mx_{ss}}, \overrightarrow{Mz_{ss}})]$
5: $Transverse\_Plane \leftarrow [polyFit(\overrightarrow{Mux_{ss}}, \overrightarrow{Mwy_{ss}}), polyFit(\overrightarrow{Mx_{ss}}, \overrightarrow{My_{ss}})]$
6: $Y \leftarrow [Sagittal\_Plane, Transverse\_Plane]$
7: **return** $Y$

Note:
[a]. For simplicity, the virtual sweet-spot marker was calculated as equidistant from the racquet's markers $\{\overrightarrow{M_{R1}}, \overrightarrow{M_{R2}}, \overrightarrow{M_H}\}$.
[b]. Gradient vector at the specific 3D point: $(\nabla f)_{M_{ss}} = grad(M_{ss})$.
[c]. Second-degree polynomial curve fitting: $[p_2, p_1, p_0] = polyFit()$.

(side and top views). Spatial patterns converted from curvature shapes are provided as input to the *Radial Basis Function* (RBF) connectionist system for classification purposes.

### D. Visualisation of Intermediate Results

The following figures show evidence of computational steps involved in the feature extraction algorithm (Table I – Algorithm 1). Fig. 4 shows a swing with the gradient vector flow originating from the computed virtual sweet spot of the racquet.

a)
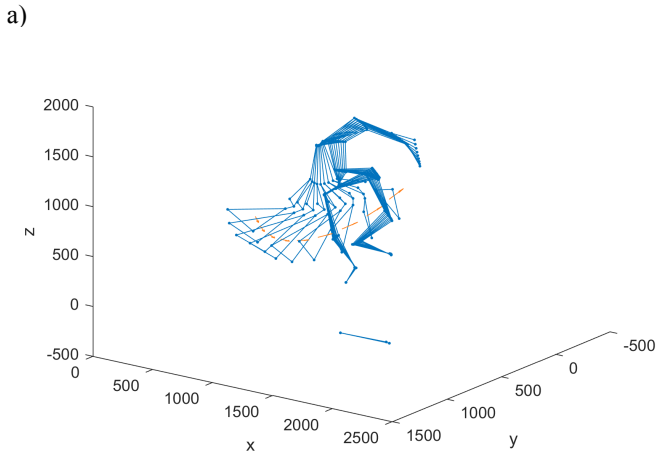

b)
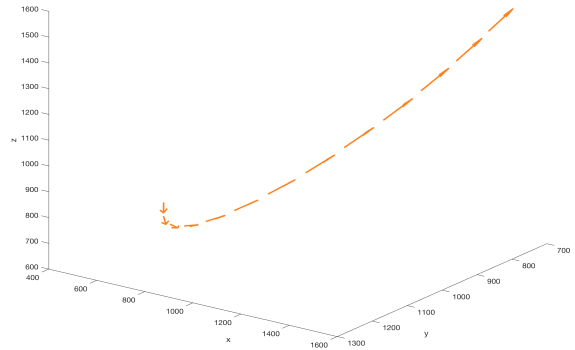

Fig. 4. Gradient vector flow of the racquet's sweet spot with (a) and without (b) player's stick figure visualisation.

The curved shapes of the sweet spot trajectory through the action zone were transformed using polynomial interpolation (7), where the polynomial parameters become variables or features.

$$f(x) = p_n x^n + p_{n-1} x^{n-1} + , ..., + p_1 x + p_0 \quad (7)$$

Visual inspection of one of the non-linear marker trajectories of the racquet (Fig. 5) and common knowledge of the racquets' mass (typically over 300 g) suggests that using a second-degree polynomial is the best option for this curve fitting model.

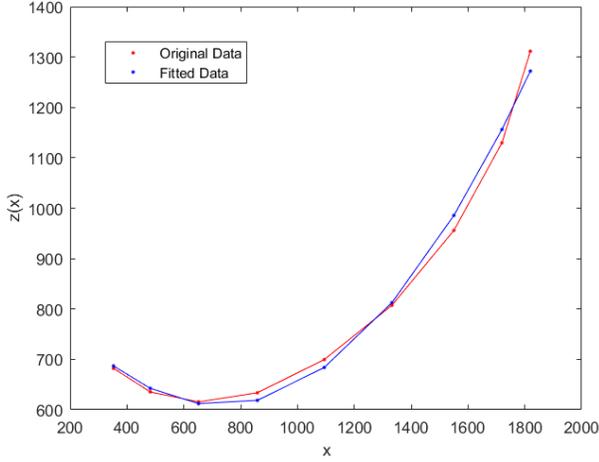

Fig. 5. Example of curve fitting model for the racquet's marker trajectory within swing's action zone on the sagittal plane (side view).

The shape of the curve (Fig. 5) shows how the racquet's string bed is producing a forward motion (direction x: left to right) and top-spin rotation (direction z(x): low to high) throughout the impact zone. Top-spin vs. ball speed depends on a player's court position but also on their personal style of play.

The benefits of the employed polynomial curve fitting approach include:

- Temporal to spatial pattern transformation
- Data compression
- High frequency noise filtering (data smoothing), which may be useful for subsequent analysis (e.g. acceleration and up-sampling).

The produced output vectors $Y$ (Table I – Algorithm 1) were linearly normalised between (-0.8 … 0.8) for RBF classifier input. Such single-marker computation is considered relatively fast, computationally inexpensive, simple and sufficient for feature extraction representing the racquet motion through the intended ball impact and swing action zone.

The chosen RBF classifier model is intended for fast classification operation requiring low computational resources. Along with other traditional ANN models (e.g. SVM and MLP), RBF is typically used for benchmarking purposes. Unlike MLP and more recent deep learning ANNs (also requiring larger training datasets), RBF processing architecture that has only one hidden layer. For future advancements, the traditional RBF is still considered a good candidate for modifications that would allow adaptive and evolving operation for incremental learning such as [8, 9]. Future advancements are likely to investigate the underlying KNN responsible for multivariate Gaussian parameter settings from data that could be modified with the evolving clustering function (ECF) or other adaptive and evolving classification model alternatives [10].

## IV. RESULTS

Capturing tacit expert assessment via supervised learning into a computer model has practical applications for the next generation of inertial sensors, wearables and optoelectronic systems. With a small training dataset, such systems could quantify the number of errors during the recorded training session for many sports disciplines beyond tennis. The key evidence of feature extraction concepts was provided as intermediate results while the performance of the produced RBF model provided the insight into flexible assessment criteria for the motion dataset.

The developed feature extraction technique (Table I – Algorithm 1), for spatiotemporal data transformation into spatial patterns is robust to the varying durations of (visually) selected tennis swings' action zone, and for each swing it produces a vector consisting of 12 variables.

Table II shows the achieved efficiency of the problem space dimensionality reduction by Motion Gradient Vector Flow (Table I – Algorithm 1).

TABLE II. SUMMARY OF ALGORITHM 1 INPUT SPACE DIMENSIONALITY REDUCTION

| Swing's ROI duration | Input Dimensionality | Output Dimensionality | Space reduction |
|---|---|---|---|
| 13 | 858 | 12 | 98.6% |
| 10 | 660 | 12 | 98.2% |
| 7 | 462 | 12 | 97.4% |

Where, for example, the selected swing's ROI duration is: 10 samples x 66-time series = 660.

Algorithm 1 (Table I) output $Y$ is: 3 (parameters/curve fitting) x (sagittal plane + transverse plane) x (sweet spot virtual marker trajectory + top of gradient vectors' curve) = 12.

Resulting in: [1-(12/660)] x 100% = 98.2% data reduction.

Regarding expert labelling and decision boundaries related to whether the observed swing was good or bad for the skill level of the player, the small dataset also reduced the time required to manually produce multiple assessments for novice and advanced players. Given the challenge posed by the small-sized dataset and random initialisation of RBF model, the LOO cross-validation was repeated 12 times and the mean classification *Accuracy* (8) was reported in Table III. Given the small dataset, the results (Table III) include sub-optimal RBF solutions to indicate possible overfitting and model performance with s sub-optimal number of hidden-layer's processing units.

The classification accuracy for each LOO cross-validation was calculated as:

$$Accuracy = 1 - \left(\frac{\sum_{i=1}^{N}\varepsilon}{N}\right) \cdot 100\% \qquad (8)$$

Where:
$N$ … the number of input vectors, and
$\varepsilon$ … is 1 for a misclassified input vector, or 0 otherwise.

Table III shows for optimal and sub-optimal RBF modelling solutions. All processing units (artificial neurons) of RBF models used in experiments have a Gaussian activation function, and model training was based on KNN clustering.

TABLE III. FLEXIBLE SKILL-LEVEL ASSESSMENT CRITERIA AND CLASSIFICATION RESULTS USING RBF CLASSIFIER MODEL

| Repeated Leave-one-out Cross-validations for Two Assessment Criteria (Novice and Intermediate) using RBF Classifier Model | | | |
|---|---|---|---|
| | Number of RBF training epochs: 100 | | |
| | Number of input vectors: 14 Number of repeated LOO cross-validations: 12 | | |
| *Observation* | *Number of RBF Hidden –Layer Processing Units* | *Novice Skill Assessment* Bad swings portion: 28.6 % | *Intermediate Skill Assessment* Bad swings portion: 71.4 % |
| Sub-optimal solution | 2 | 71.4% | 85.7% |
| | 3 | 83.3% | 91.7% |
| Optimal | **4** | **84.5%** | **94.6%**[a] |
| Potential overfitting | 5 | 85.1% | 93.4% |
| | 6+ | N/A[b] | N/A[b] |

[a.] Intended RBF model would converge with minimal epochs (5-12) out of 100 training epochs limit.
[b.] Reported errors – where the RBF model training may not converge towards the intended solution.

The results (Table III) show the difference (approx. 10%) in classification accuracy for diverse skill assessment training data on the same motion dataset. Better classification accuracy for the intermediate skill level than for novices, suggests that more follow-up research is needed to investigate potential RBF model 'awareness' on single and compound technique errors found in 'bad' swings compared to more consistent 'good' swings.

V. DISCUSSION

The human body and racquet was modelled as a set of interconnected rigid segments that should be sufficient for visual analysis of human motion without the potential for human bias (e.g. potentially caused by liking/disliking a player, prior knowledge or recent observation of a player, clothing, height, body shape, or gender). As part of our motion perception, the majority of people are able to sense whether a movement is 'natural' or not, demonstrated, for example, by the ability to distinguish between animations created by animation artists only and those created by using motion capture. For such reasons, the experimental design did not include synthetic data or attempts to reconstruct incomplete marker trajectories needed to compute a virtual sweet spot from the captured motion dataset. Further swing reduction from the captured motion data was limited to forehands only. Visual examinations of the captured forehand swings have shown better coverage and variations than the backhand swings (e.g. stance coverage, swing width, top-spin variations and swing durations). In addition, for novice players, forehand is typically easier to learn than the backhand (whose learning may progress at different pace than forehand). The high classification accuracy produced by using a relatively small dataset for modelling purposes provides an advantage for practical applications, such as where a coach would like to automate the tagging of erroneous or good shots for analytical replay purposes or for the next generation of exergames. Furthermore, it would be beneficial to initially use small datasets and later to employ adaptive and incremental learning capabilities that would still rely on occasional human expert labelling.

One of the obstacles for research rigour was stick figure 3D replay for expert visual assessment, consistent selection of action zone ROI for swings of varying durations, and consistent data labelling based on qualitative analysis of human movement and coaching practise. As some swings were harder to assess than others, a standalone 3D player was developed with a proprietary graphics library that provided smooth and pixel-accurate interactive virtual camera movement during the replay [7]. For the research community using Matlab, Octave or similar, the 3D stick figure player code can be implemented using `plot()` or `plot3()` functions with fewer lines of code than if implemented in C++, Java or Object Pascal (Delphi or Lazarus). For computational sport science, HMMA and expert labelling video and 3D replay tools are considered essential, since feature space and internal workings of an ANN are typically not comprehensive for human learning or understanding. Extended functionality for augmented coaching using video and 3D stick figure replay was reported in [11].

Stick figure replay and silhouette filtering [12] may be used for coaching, on-line coaching and also encourage participation in on-line exergaming by facilitating privacy preservation, such as that needed for healthcare/elderly-care monitoring systems.

The presented novel approach and ideas were driven to support model design that can operate on initially small to large datasets and for spatiotemporal motion patterns for which there are no statistical ground truth available. Using relatively small training data from the coach, the aim was to solve the 'curse of dimensionality' associated with kinematic 3D data analysis of diverse forehand swings [13]. More and less rigid criteria for swing technique assessments for novices and intermediate skill levels reflect subjective and qualitative nature of coaching practise, where feedback is focused on performance elements rather than on the knowledge of the outcome, that is ball flight.

Single virtual marker computation concepts are transferrable for practical use with for example, smart watch inertial sensors or an inertial sensor attached to the racquet or other sports equipment. Using computer vision (video or depth video), it would be possible to combine or fuse data from diverse sources to enable advancements of HMMA with a high sampling frequency around ROI and video replay for the next generation of ACST and exergames.

VI. CONCLUSION

The next generation of exergames, and augmented coaching/rehabilitation systems and technology are expected to provide analytical capabilities that will help an end user to improve his/her sport-specific technique or (re)gain motor skills. This multidisciplinary paper presents concepts and solutions that viably automate the classification of good or bad movement patterns based on qualitative, flexible and subjective criteria similar to a coach.

The presented artificial neural network-based solution can, in part, mimic a coach, who can immediately tell if the observed swing (or other movement pattern) does not 'feel right' before providing subsequent analysis with descriptive/qualitative feedback to improve element(s) of performance. Relying on 3D stick figure replay of the recorded dataset, it was possible to capture two different subjective assessment criteria. The use of two different assessment criteria reflects expected swing techniques for novices and more advanced skill-level players. The achieved classification of tennis swings (with accuracy of: 84.5% for novices and 94.6% for intermediate-level players) demonstrates a flexible and personalised machine learning solution for designing exergaming and augmented coaching systems and technology.

The presented mathematical transformation concepts involved in the presented feature extraction technique, insights and neural data processing techniques have practical aspects that could be transferred to a number of sport disciplines or rehabilitation scenarios. Potential examples include: (i) keeping track of person-specific sub-standard movements at the end of a race, game or training session – a system could quantify a number of motion patterns executed with poor technique using small initial system-training data from a coach who is familiar with the player's idiosyncrasies; (ii) autonomously tagging irregular spatiotemporal patterns for replay purposes; (iii) recording intellectual property into a machine that could capture an expert's tacit knowledge by combining swing replays (or other sport-specific motion patterns) with output labelling; (iv) healthcare monitoring, to automate finding of irregular signal patterns from logging devices (e.g. a holter cardiac monitoring system and other auscultation expert systems); and (v) an augmented coaching system to supervise, in a controlled environment, a patient's activities by monitoring for erroneous movement patterns that could adversely affect rehabilitation time.

The concepts and technique presented in this paper utilised a motion dataset that was captured at 50 Hz without ball impact information. Considering Microsoft Kinect™ sensor's streaming capabilities (streaming at 30 Hz), recent mobile and sport camera video capabilities (mono and stereo vision at 120 and 240 Hz) and the ability of some inertial sensors to capture motion data close to 1000 Hz, it is likely that this work will be compatible with further motion capture technology advancements. The concepts of the racquet's feel and sweet spot expressed as single virtual marker data processing techniques are generally applicable to neural information processing and utilisation of data fusion from diverse sources (e.g. sport equipment-attached, smartwatch or other wearable sensors with computer vision).

Future work will be focused on advancements of machine learning approaches for technique assessment for augmented coaching systems, wearables, and rehabilitation devices. Another broader avenue to pursue is computational sport science that will also include the implementation of evolving and adaptive systems, deep learning and the third generation of ANN to advance human motion modelling and analysis applicable to human motor learning, skill and technique (re) acquisition, and knowledge-discovery from diverse disciplines datasets.

ACKNOWLEDGEMENTS

The author wishes to thank Prof. Ian Nabney and Dr. Christopher Bishop for sharing and updating their source code for Netlab and RBF for the past two decades. This multidisciplinary work was also supported by New Zealand tennis coaches Shelley Bryce (née Stephens) and Kevin Woolcott who took their time and efforts by participating in 3D replay assessment of the tennis dataset utilised in this study.

The dataset has been recorded in Peharec Polyclinic (Pula, Croatia) with help of MSc. Petar Bačić, who provided valuable multidisciplinary assistance with motion capture, biomechanics support and tennis coaching critiques during data capture. Free laboratory access and resource-sharing was granted by Dr. Stanislav Peharec.